\documentclass[a4paper,10pt]{article}
\usepackage[latin1]{inputenc}
\usepackage[T1]{fontenc}
\usepackage{natbib}
\bibpunct{[}{]}{;}{a}{,}{,}
\usepackage[normalem]{ulem}
\usepackage{verbatim}
\usepackage{amssymb}
\usepackage{vmargin}
\usepackage{graphicx}
\setmarginsrb{2.5cm}{2.5cm}{2.5cm}{2.5cm}{0cm}{0cm}{0cm}{1cm}
\title{\textbf{Optohydrodynamics of soft fluid interfaces: optical and viscous nonlinear effects}}
\author{By\\ Hamza CHRAIBI$^{1,2}$, Didier LASSEUX$^1$, R\'egis WUNENBURGER$^2$,\\ Eric ARQUIS$^1$ and Jean-Pierre DELVILLE$^2$\\
\\ 
 $^ 1$ \small Universit\'e Bordeaux I, TREFLE (UMR CNRS 8508)\\\small Esplanade des Arts et M\'etiers, 33405 Talence Cedex,
France.\\$^ 2$ \small Universit\'e Bordeaux I, CPMOH (UMR CNRS 5798)\\ \small351 cours de la Lib\'eration, 33405 Talence cedex, France.}

\begin{document}

\maketitle

\begin{abstract}
\large
Recent experimental developments showed that the use of the radiation
pressure, induced by a continuous laser wave, to control fluid-fluid
interface deformations at the microscale, represents a very promising
alternative to electric or magnetic actuation. In this article, we 
solve numerically the dynamics and steady state of the fluid interface under the effects of buoyancy, capillarity, 
optical radiation pressure and viscous stress.
A precise quantitative validation is shown by comparison with experimental data. New results due to the nonlinear dependence of the optical pressure on the angle of incidence are presented, showing different morphologies of the deformed interface going from needle-like to finger-like shapes, depending on the refractive index contrast. In the transient regime, we show that the viscosity ratio influences the time taken for the deformation to reach steady state.

\end{abstract}

\section{Introduction}
\begin{figure}[h!!!]
\begin{center}
 \includegraphics[scale=0.4]{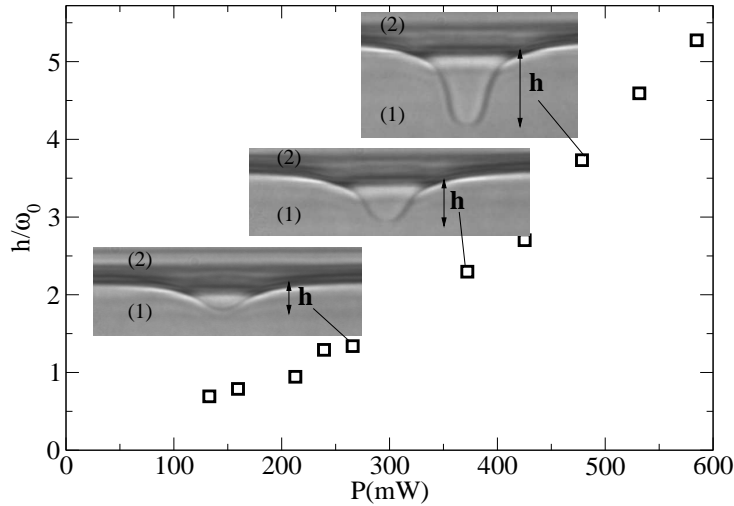}
\caption{Nonlinear variation of the height $h$ (scaled by the beam radius $\omega_0=7.5 \mu m$) of the deformed interface versus the incident beam power $P$. The laser wave propagates upwards. The pictures corresponds to $P=265,372$ and $478 mW$. Interfacial tension is $\sigma=5~10^{-7} N/m$, refractive index contrast is $n_2-n_1=0.012$ and density contrast is $\rho_1-\rho_2=24.7~ kg/m^3$.}\label{exp}
\end{center}
\end{figure}

\large
Optical waves are an interesting and promising alternative to pure electric %
\citep{wohlhuter92} and magnetic fields \citep{sherwood87} to produce mechanical stress on fluid interfaces and
deform them.\\ When focused light propagates through an interface separating two fluids of
different refractive indices, photon momentum, which linearly depends on the
refractive index, experiences a jump at the interface. This momentum jump
induces a radiation pressure on the interface that acts towards the fluid
of lowest optical index whatever the direction of propagation of the optical
wave.\\ This surprising invariance property was first evidenced in the experiments of Ashkin and Dziedzic \citep{ashkin73}. Their work was an attempt at giving an experimental answer to the Abraham-Minkowski controversy about light momentum's expression in a dielectric. The controversy arises due to Abraham's and Minkowski's predictions disagreeing as to whether the momentum carried by an electromagnetic field is increased (Minkowski) or
decreased (Abraham) by the presence of a refractive medium. Ashkin and Dziedzic concluded that light momentum was consistant with the Minkowski formalism (see \citep{pfeifer07} for a recent review on light momentum).\\ The deflection of the interface towards the less refractive fluid was later confirmed by Zhang \& Chang \citep{zhang88} who used a pulsed laser
wave to deform a water droplet and observed, immediately after the pulse, an
oscillation of the interface leading to an emission of
microdroplets at the rear of the drop at large pulse energy. More recently, Casner \& Delville %
\citep{casner01a} used the soft interface
separating two near-critical liquid phases in order to reduce by several orders
of magnitude the effect of interfacial tension and thus 
to significantly enhance the amplitude of interface deformations. Stationary deformations of large aspect ratio were observed \citep{casner03b} as illustrated in Figure \ref{exp}.\\

The radiation pressure effects open interesting perspectives
for applications in many different fields: \\
(i) interface rheology first, with
measurements of surface and interfacial tension at small scale %
\citep{sakai01,mitani02}, and microrheology characterization \citep{yoshitake05,wottawah05}\\ (ii) adaptive optics and holography with light actuation of reconfigurable fluid
lenses \citep{ashkin73,kats79,casner01b} and interface relief gratings %
\citep{komissarova88}\\ (iii) droplet cavity lasing \citep{hartings97}\\
(iv) surface relief micropatterning by extending electrical field
approaches \citep{schaffer00} to optics. 

While experiments have highlighted a rich phenomenology, the complete
understanding of the physics of interface deformation induced by an optical
wave is still in its infancy. This task is
indeed made difficult by the coupling
 between the shape of the interface and the optical radiation pressure
distribution as well as the dependence of the interface shape on buoyancy, capillarity, viscosity and optical properties of the fluids. While some effort has been dedicated to the description of buoyancy
effects on equilibrium interface deformations of small amplitude \citep{wunenburger06a} (i.e. linear regime where the deformation's height varies linearly with the beam power), very few studies have gone one to describe large amplitude deformations %
\citep{hallanger05,chraibi08a}.\\
The aim of this paper is to investigate numerically the statics and dynamics of these large deformations for a wide range of refractive index contrasts, from very soft interfaces (near-critical systems $\sigma\sim10^{-4}mN/m$) to more usual pairs of fluids (water/air for instance, $\sigma\sim10-100mN/m$ ). 
In essence, it is shown that the underlying physics can be explained by the nonlinear dependence of the radiation pressure on Snell-Descartes angles of refraction.
It is further shown that the transient shape of the deformation exhibits a transition at characteristic time which depends on the viscosity ratio.\\
Our analysis is organized as follows. The physical model for interface deformation is presented in section 2. In section 3, we first validate the computation algorithm
through a successful comparison with experimental results. Finally an analysis of the refractive index contrast and viscous effects is performed showing new important features in both the unsteady and steady states in the nonlinear regime of deformation, i.e. when the equilibrium hump height has a nonlinear variation with the excitation amplitude. 

\section{Governing equations and numerical resolution}
\subsection{Physical model}
\begin{figure}[h!!!]
\begin{center}
 \includegraphics[scale=0.55]{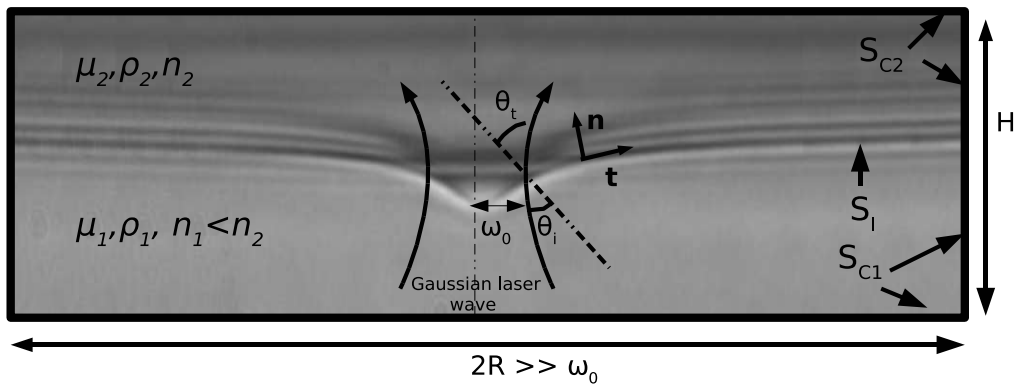}
\caption{Configuration and notations considered for simulations illustrated using an experimental liquid-liquid interface deformation driven by a focused laser
wave propagating upward. $n_2-n_1=0.011$, $\rho_1-\rho_2=22~kg/m ^3$, $\sigma=3.6~10^ {-7}N/m$; $\omega_0=5.3 \mu m $ ;
$P=240 mW$. $S_I$, $S_{C1}$ and $S_{C2}$ respectively denote the interface and the solid walls in fluids $1$ and $2$.}\label{sketch}
\end{center}
\end{figure}

We investigate the deformation of a initially horizontal stationary liquid-liquid interface, induced by the optical radiation pressure due to a continuous Gaussian laser beam of power $P$ and radius (also called beam waist) $\omega_0$. Physical properties of the liquids (denoted $1$ for the bottom and $2$ for the top) are their refractive indices $n_1$, $n_2$, viscosities $\mu_1$, $\mu_2$ and densities $\rho_1$, $\rho_2$. The interfacial tension is denoted by $\sigma$.
 The fluids are enclosed in a cylindrical cell of radius $R\gg\omega_0$ and total height $H\gg\omega_0$.\\
Considering axisymmetry along the beam propagation axis $z$, cylindrical coordinates (${\bf e_r, e_\alpha, e_z})$ with origin $O$ located at the intersection of the beam axis
with the initial flat interface are chosen for this study.  A point $\textbf{x}$ is thus referenced by the space coordinates $(r,\alpha,z)$. The configuration is summarized in figure \ref{sketch}. \\

\subsubsection{The optical radiation pressure}
We consider the fluids as two non-absorbing dielectric media with different refractive indices $n$, separated by an interface of arbitrary
shape. The indices $i$ and $t$ refer to incidence and transmission, and $\theta_i$ and $\theta_t$ are,
respectively, the Snell incident and transmission angles (see figure \ref{sketch}). They can be expressed as $\theta_i=\arctan(\frac{d z}{dr})$ and $\theta_t=\arcsin((1-\delta) \sin\theta_i)$ where $\delta=\frac{n_2-n_1}{n_2}$ is the relative index contrast and $z=z(r)$ is the interface equation. Since the photon momentum depends on the refractive index, light momentum is not conserved when the beam crosses the interface separating these
two dielectrics. The resulting discontinuity in momentum
gives birth to the radiation pressure applied to the interface. If ${\bf t}$ and ${\bf n}$ refer to the tangent and normal directions to the interface at the location where the light impinges the interface, $\nu_0$ represents the optical wave frequency, $h_0$ the Planck constant and $c$ the light speed in vacuum, the photon momentum balance can be deduced from the fact that :\\ (i) an
incident photon gives the momentum $(n_1 h_0 \nu_0 /c)(\sin\theta_i{\bf t}+\cos \theta_i {\bf n})$ to the interface, (ii) a reflected photon carries the momentum $(n_1 h_0 \nu_0 /c)(\sin\theta_i{\bf t}-\cos \theta_i {\bf n})$ away from the interface and (iii) a transmitted photon picks the momentum $(n_2 h_0 \nu_0 /c)(\sin\theta_t{\bf t}+\cos \theta_t {\bf n})$ to the interface. By denoting $N$ the
number of photons impinging on the interface per unit time and
unit interfacial area and $R(\theta_i,\theta_t)$ and $T(\theta_i,\theta_t)=1-R(\theta_i,\theta_t)$ the classical Fresnel coefficients of reflection and transmission of electromagnetic energy for circularly polarized beams are given by \citep{stratton41}:
\begin{equation}
T(\theta _{i},\theta _{t})=\frac{2(1-\delta)\cos \theta _{i}\cos \theta
_{t}}{((1-\delta)\cos \theta _{i}+\cos \theta _{t})^{2}}+\frac{%
2(1-\delta)\cos \theta _{i}\cos \theta _{t}}{(\cos \theta
_{i}+(1-\delta)\cos \theta _{t})^{2}}.
\end{equation}
 We can express the momentum
variation $d{\bf Q}$ of the incident beam on an interface element of area $dS$ during the time $dt$ as  $d{\bf Q}=d{\bf Q_t}+d{\bf Q_n}$, i.e.:
\begin{eqnarray}
\lefteqn{d{\bf Q}=d{\bf Q_t}+d{\bf Q_n}=[n_1 \sin\theta_i-(Rn_1 \sin\theta_i+T n_2 \sin\theta_t)]\frac{Nh_0\nu_0}{c}dSdt{\bf t}}\nonumber\\
&& +[n_1 \cos\theta_i-(-Rn_1 \cos\theta_i+T n_2 \cos\theta_t)]\frac{Nh_0\nu_0}{c}dSdt{\bf n} 
\end{eqnarray}
Since $n_1 \sin\theta_i=n_2 \sin\theta_t$ then $d{\bf Q_t}=0$. There is no momentum transfer
parallel to the interface. Consequently, one has
\begin{equation}
d{\bf Q}=d{\bf Q_n}=n_1 \cos\theta_i\left(1+R-\frac{\tan\theta_i}{\tan\theta_t}T\right)\frac{Nh_0\nu_0}{c}dSdt{\bf n}.
\end{equation}
Classically, the laser intensity $I$ is defined as $I=N_0h_0\nu_0$, where
$N_0$ is the flux of photons through the beam section. Since the
laser wave incidence angle on the interface is $\theta_i$, one gets $N=N_0 \cos\theta_i$. We deduce that the optical radiation force acting on the interface per unit area $dS$, when the laser wave propagates from the less to the more refractive fluid, ${\bf \Pi^ {-+}}$, is normal to the interface and given by
\begin{equation}
{\bf \Pi^ {-+}}(r)=n_1 \cos^ 2\theta_i\left(1+R-\frac{\tan\theta_i}{\tan\theta_t}T\right)\frac{I(r,z)}{c}{\bf n}\label{radup}
\end{equation}
where ${\Pi^ {-+}}$ is the radiation pressure.\\
When the beam is weakly focused, the $z$ dependence of the beam radius can be neglected, yielding :
\begin{equation}
I(r,z)\approx I(r)=\frac{2P}{\pi \omega _{0}^{2}}e^{-2(r/\omega_0)^{2}}.
\end{equation}
\begin{figure}[h!!!]
\begin{center}
 \includegraphics[scale=0.8]{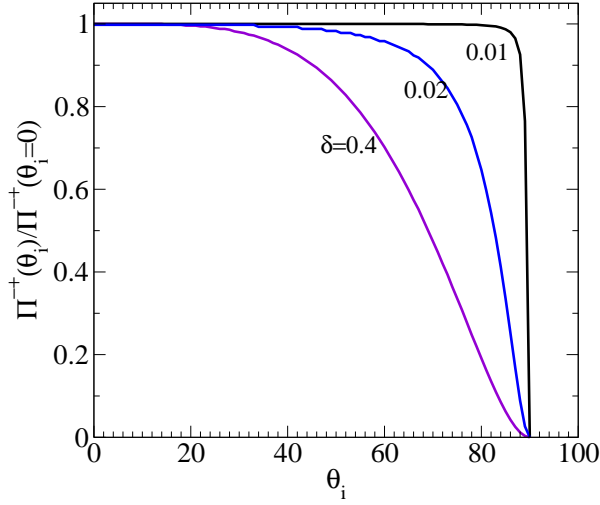}
\caption{Variation of the radiation pressure normalized by its
value at normal incidence, versus the incidence angle for different values of the relative index contrast $\delta=\frac{n_2-n_1}{n_2}$.}\label{pi}
\end{center}
\end{figure}
In figure \ref{pi}, we have represented $\Pi^{-+}(\theta_i)$ normalized by $\Pi^{-+}(\theta_i=0)$, the radiation pressure at normal incidence, versus the incidence angle $\theta_i$ for different values of the relative index contrast $\delta=\frac{n_2-n_1}{n_2}=0.01, 0.02$ and $0.4$.\\ First, we can see that the normalized radiation pressure decreases with $\theta_i$ and vanishes at tangential incidence ($\theta_i=90°$). Second, $\Pi^{-+}(\theta_i)$ is independent of the incidence angle when $\theta_i<20°$, a situation that corresponds to the regime of quasi-normal incidence. For $\theta_i>20°$, $\Pi^{-+}(\theta_i)$ decreases significantly beyond a certain value of $\theta_i$ which increases as $\delta$ decreases. Physically, this illustrates that the radiation pressure, i.e. the momentum jump of photons at the interface, is less sensitive to the incidence angle when the two fluids tend to refractive index matching ($\delta\rightarrow0$). As detailed below, the coupling between radiation pressure and interface deformation ($\tan \theta_i=\frac{dz}{dr}$), is the major physical feature that explains both the dynamics and statics of the interface deformation.\\ 

\subsubsection{Governing equations}
Since the Reynolds number associated with the flows under consideration is always small compared to unity ($Re=\frac{U \omega_0 \rho}{\mu}\simeq 10^ {-3}$, $\mu\simeq 10^{-3}Pa.s$, $\rho\simeq10^ 3kg/m^ 3$, $\omega_0\simeq10\mu m$ and $U\simeq10^{-4}m/s$ a characteristic velocity of the flow), both fluids obey the mass conservation and Stokes equations.\\
 Their dynamics are coupled through stress balance in addition to the continuity of velocity at the interface denoted $S_I$. The motion of the interface is described following a Lagrangian approach. Using $w_0$ as a reference length,  $\displaystyle{U_0=\frac{2\sigma}{\mu_1+\mu_2}}$  as a reference velocity and $p_{i0}=\frac{\mu_i U_0} {\omega_0}$~,~$i=1,2$ as a reference pressure in fluid $i$, the axisymmetric dimensionless boundary value problem can thus be expressed as\\
\begin{equation}
\nabla . {\bf u_i} = 0~~;~~  i=1,2
\label{mass}
\end{equation}
\begin{equation}
{\bf 0} = -\nabla p_i + \Delta {\bf u_i}~~;~~ i=1,2
\label{stokes}
\end{equation}
\begin{equation}
\frac{2}{1+\zeta}(\zeta {\bf T_{1}}\cdot {\bf n}-{\bf T_{2}}\cdot {\bf n})=(\kappa (r)+\Pi (r) - Bo ~ z )
{\bf n}~~;~~ {\bf x}~ on~ S_I \label{stressjump}
\end{equation}
\begin{equation}
{\bf u_1} = {\bf u_2}~ on~ S_I~;~{\bf u_1=0}~on~S_{C1}~;~{\bf u_2=0}~on~S_{C2}
\end{equation}
\begin{equation}
\frac{d{\bf x}}{dt}={\bf u}({\bf x}){ ~~~~ }on~ S_{I}
\label{kin}
\end{equation}%
${\bf u_i}$, $p_i$ are respectively the dimensionless velocity and pressure in each fluid while \\$\displaystyle{{\bf T_i}= -p_{i} {\bf I} + (\nabla {\bf u_i} + ^t\nabla {\bf u_i})}$  is the dimensionless stress tensor and $\zeta = \mu_1/\mu_2$ the viscosity ratio. $\displaystyle{\kappa(r)=\frac{1}{r}
\frac{d}{dr}\frac{r\frac{dz}{dr}}{\sqrt{1+{\frac{dz}{dr}}^2}}}$ is twice the mean curvature of the
axisymmetric interface in cylindrical coordinates, $\Pi(r)=\frac{\omega_0}{\sigma}\Pi^ {-+}(r)$ is the dimensionless radiation pressure where $\Pi^ {-+}$ is given by equation (\ref{radup}) and $Bo=(\rho_1-\rho_2) g \omega_0^2 / \sigma$ is the gravitational Bond number which quantifies the relative magnitude of buoyancy to capillary effects for small deformation amplitude.
A detailed investigation dedicated to the gravitational effects, was proposed in a previous work, for small deformation amplitudes \citep{wunenburger06a,wunenburger06b}. Since $Bo$ is typically in the range $10^{-3}$..$10^{-1}$ for fluids commonly encountered, buoyancy can reasonably be neglected compared to the other forces and will not be discussed in the present paper.\\
In order to quantify effects of the laser wave on the interface deformation,
it is convenient to define the optical to Laplace pressure ratio.\ This ratio  $%
\pi_0 $ taken at $r=0$ ($\theta _{i}=\theta _{t}=0)$, is defined as:
\begin{equation}
\pi_0= |\Pi {(r=0)}| =\frac{4P}{\pi c\omega _{0}\sigma }%
\frac{n_1 (  n_{2}-n_{1})}{(n_{2}+n_{1})}~
\end{equation}
and can be considered as the optical Bond number.\\

\subsection{Integral formulation and algorithm}

Since solutions to the Stokes problem can be formulated in terms
of Green's functions \citep{pozrikidis92}, we rewrite the
governing equations as a system of integral equations over the
boundaries of the computational domain. Once boundary conditions
on the interface $S_I$ and solid boundaries $S_{C1}$ and $S_{C2}$ in contact with fluid 1 and 2 are used (see Figure \ref{sketch}), the two-phase Stokes
problem can be written in the following compact form:
\begin{eqnarray}
\lefteqn{\frac{1+\zeta}{2}{\bf u(x)} = \int_{S_I} {\bf
U.n}(\kappa(r_y)+\Pi(r_y)-Bo~z(r_y))dS_y +} \label{zebigone}\\
& & (1-\zeta)\int_{S_I}{\bf n.K.}{\bf
u}dS_y+\zeta\int_{S_{C1}}{\bf U.(T_1.n)}dS_y-\int_{S_{C2}} {\bf
U. (T_2.n)}dS_y. \nonumber 
\end{eqnarray}

Here, \textbf{U} and \textbf{K} are Green's kernels for velocity and
stress respectively and are given by \citep{pozrikidis92}:
\begin{eqnarray}
{\bf U(d)}&=&\frac{1}{8\pi}(\frac{1}{d}{\bf I}+\frac{{\bf dd}}{d^3}),\\
{\bf K(d)}&=&-\frac{3}{4\pi}(\frac{{\bf ddd}}{d^5}),
\end{eqnarray}
where ${\bf d=x-y}$, ${\bf y}(r_y,z_y)$ is the integration point.
In Eq. (\ref{zebigone}), the first term in the right hand side
describes the flow contribution from interfacial tension,
radiation pressure and gravity, whereas the second term accounts
for shear rates contrast on the interface. This term vanishes when
$\zeta=1$. The third and fourth terms account for shear
occurring on $S_{C1}$ and $S_{C2}$ as a result of the no-slip
boundary condition.

Velocities on the interface as well as stress over all the
boundaries $S_{I}$, $S_{C1}$ and $S_{C2}$ are determined by
solving the discrete form of this equation using a Boundary Element Method (BEM). Details on the BEM applied to two-phase axisymmetric flow can be found in the review by Tanzosh et al. \citep{tanzosh92} on the solution
of free surface flow using this technique. The BEM reveals to be an excellent tool to
solve interfacial flow problems with high resolution as reported
in the analysis of flow involving electric and magnetic fields
\citep{sherwood87} or buoyancy
\citep{manga94,koch94}.\\
 The solution requires first the discretization of all
the boundaries $S_{I}$, $S_{C1}$ and $S_{C2}$. Due to the integral
formulation and axial symmetry, the problem is reduced to one
dimension and only line boundaries need to be discretized. In this work, the mesh is made of constant
boundary elements, i.e. line segments with centered nodes. Azimuthal integration
of Eq. (\ref{zebigone}) is performed analytically
\citep{lee82,graziani89} reducing Eq. (\ref{zebigone}) to a line
integration which is finally performed using Gauss quadratures
\citep{davis84}. Elliptic integrals resulting from the azimuthal
integration are evaluated using power series expansions
\citep{bakr85}.\\  The fluid-fluid interface $S_{I}$ is parameterized in terms of arc
length and is approximated by local cubic splines to remesh the interface during its deformation, so that the
curvature can be accurately computed. Distribution and number of
points are adapted to the shape of the interface, so that the
concentration of elements is higher in regions where the variation
of curvature of the interface is large. The number of mesh points
is about 70 for a typical computation of a small interface
deformation. The solid boundaries $S_{C1}$ and $S_{C2}$ are meshed
using about 40 uniformly distributed points. An increase in the
mesh resolution for the interface and the solid boundaries do not
show any change in the results.\\ The motion of the interface is followed using
the kinematic condition (\ref{kin}) which is discretized using an
explicit first-order Euler time scheme. A typical computation
begins with a flat interface at rest. The laser beam is switched
on at $t=0$, and the interface deforms towards fluid $1$ of
smallest refractive index. Computation stops when an equilibrium
state is reached ($\displaystyle{{\bf v.n}|_{S_I}\rightarrow0}$). 
The time step is chosen to be
about 20 times smaller than the reference time $\tau_0=\frac{\omega_0}{U_0}$.
%Note that the equilibrium shape of the interface was also solved using a finite difference method and an iterative shooting alogrithm. The steady state interface profiles resolved with boundary element methods showed excellent adequation with the results given by the finite difference method.
\section{Results}
\subsection{Comparison with experiments}
In this section, we compare numerical predictions to experimental data giving both the time evolution of the interface and the steady state hump heights and interface shapes for various laser illuminations and fluid properties.\\
The experimental fluids correspond to a phase-separated
near-critical water-in-oil micellar mixture contained in a thermo-regulated spectroscopic
cell at temperature $T$, for which a very low interfacial tension can be
achieved. Details on how the solution is prepared can be found in %
\citep{casner01a}. Above the critical temperature $T_{c}=308K$, two distinct
fluid phases $1$ and $2$ of different compositions coexist with densities $\rho _{1}>\rho
_{2}$ and refractive indices $n_{1}<n_{2}$. A vertical upward Gaussian beam from
a continuous wave Ar$^{+}$ laser operating at wavelength $514.5$ $nm$ is then
focused at normal incidence onto the flat fluid-fluid interface.
More exhaustive experimental details on the configuration and protocol as well as on the determination of the sample properties 
used here were reported earlier \citep{wunenburger06a,chraibi08a}.\\
As already demonstrated \citep{jeanjean88}, the thermophysical properties of the fluids can be evaluated using asymptotic scaling laws of near critical phenomena \citep{beysens82}.\\
Assuming that the coexistence curve is symmetric versus the critical concentration $\varphi_c$, the micellar concentration in each
phase $\varphi_i$ $(i=1,2)$ can be estimated by:
\begin{eqnarray}
\varphi_1&=&\varphi_c+\frac{\Delta \varphi}{2}, \\
\varphi_2&=&\varphi_c-\frac{\Delta \varphi}{2},
\end{eqnarray}
with $\varphi_c=0.11$ and:
\begin{equation}
\Delta \varphi=\Delta \varphi_c \left( \frac{T-T_c}{T_c}
\right)^{0.325},
\end{equation}
with $\Delta \varphi_c=0.5$.\\

The density of each phase $\rho_i$, $i=1,2$ can be written as a
function of $\varphi_i$:
\begin{equation}
\rho_i=\rho_{mic}\varphi_i+\rho_{cont}(1-\varphi_i),
\label{densite composite}
\end{equation}
where $\rho_{mic}=1045$~kg.m$^{-3}$ and
$\rho_{cont}=850$~kg.m$^{-3}$ are the densities of the micelles
and continuous phases respectively.\\

Because the average distance between two micelles is small compared to
the wavelength of the laser wave, the mixture can be regarded as
homogeneous from the electromagnetic point of view. Thus, the relative dielectric permittivity
$\epsilon_i=n_i^2$ of the mixture is \citep{landau60}:
\begin{equation}
\epsilon_i(\varphi_i)=\varphi_i\epsilon_{mic}+(1-\varphi_i)\epsilon_{cont}-\frac{\varphi_i(1-\varphi_i)(\epsilon_{mic}-\epsilon_{cont})^2}{3(\varphi_i\epsilon_{mic}+(1-\varphi_i)\epsilon_{cont})}.
\label{permittivite composite}
\end{equation}
along with $\epsilon_{mic}=1.86$ and $\epsilon_{cont}=2.14$,
$\epsilon_{mic}$ and $\epsilon_{cont}$ being the relative
permittivity of the micelles and the continuous oil phase respectively.\\
Since concentrations are weak, we use Einstein's
relation to estimate the dynamic viscosity $\mu_i$ of each phase:
\begin{eqnarray}
\mu_1=\mu_0\left(1+2.5\frac{\Delta \varphi}{2}\right)\\
\mu_2=\mu_0\left(1-2.5\frac{\Delta \varphi}{2}\right),
\end{eqnarray}
with $\mu_0=1.269 $~mPa.s.\\
Finally, interfacial tension is written as:
\begin{equation}
\sigma=\sigma_0\left(\frac{T-T_c}{T_c}\right)^{1.26},\label{gamma}
\end{equation}
with $\sigma_0=10^{-4}$~N/m.\\
Since fluids are assumed to be transparent (the optical absorption is $3~10^{-4}cm^{-1}$), the action of light on the interface can be considered as a
pure mechanical stress, allowing us to discard heat dissipation and
thermocapillary effects and to consider all liquid properties ($\rho _{i}$, $\mu_i$, $%
n_{i}$ and $\sigma $) as constant and independent of field strength.\\

Typical values of the fluids properties are given in table \ref{table}.

\begin{table}[h!!!]

\begin{center}
\begin{tabular}{|*{2}{c|}c|c| c|r|}
   \hline
   $T-T_c(K)$ &$\sigma(mN/m)$& $n_2-n_1$ & $\delta$ & $\rho_1-\rho_2(Kg/m^ 3)$&$\mu_1-\mu_2(Pa.s)$ \\
   2 &$1.8~10^{-4}$ &0.0097 & 0.0066 & 19 & $3.1~10^ {-4}$\\
   4 &$4.0~10^{-4}$ &0.012 & 0.0083 &24 & $3.9~10^ {-4}$\\
   6 &$7.0~10^{-4}$ &0.014 & 0.0095 &27 & $4.4~10^ {-4}$\\
   8 &$1.0~10^{-3}$ &0.015 & 0.010 &30 & $4.9~10^ {-4}$\\
   10 &$1.3~10^{-3}$ &0.016 &0.011  &32 & $5.2~10^ {-4}$\\
   15 & $2.2~10^{-3}$&0.018 & 0.013 & 36&$6.0~10^ {-4}$ \\
   20 &$3.2~10^{-3}$ &0.020 & 0.014 & 40&$6.5~10^ {-4}$ \\
   \hline
\end{tabular}
\caption{Typical values of fluid properties for different values of $T-T_c$}\label{table}
\end{center}
\end{table}
\subsubsection{Unsteady regime}
\begin{figure}[h!!!]
\begin{center}
 \includegraphics[scale=0.55]{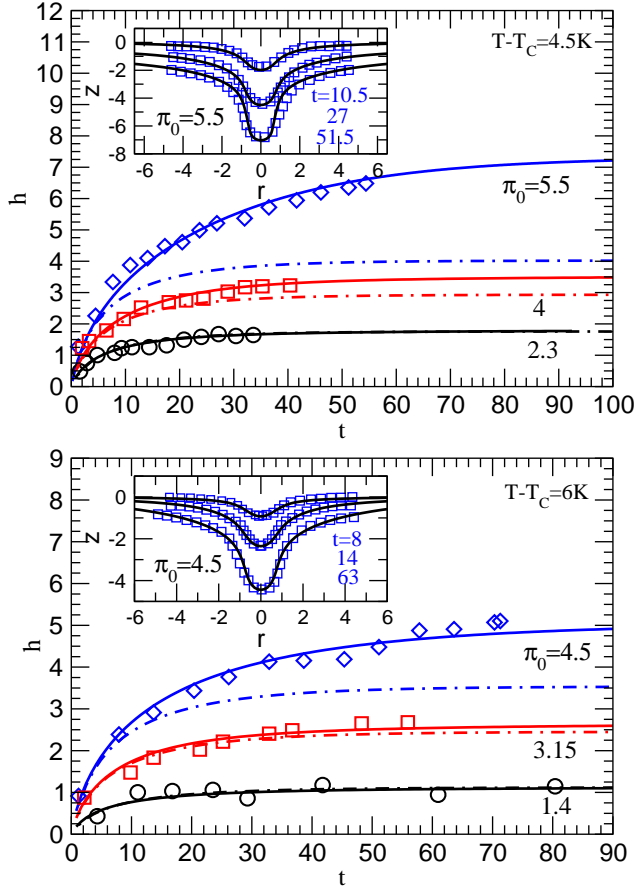}
\caption{Time evolution of the interface hump height for $\omega_0=4.8 \mu m$. Experiments (symbols) are compared to numerical predictions (continuous lines) and linearized theory $h_{lin}(t)$ (dot-dashed lines).
 (Top)$T-T_c=4.5K$, $\delta=\frac{n_2-n_1}{n_2}=0.0086$ and $\zeta=\frac{\mu_1}{\mu_2}=1.37 $. (Bottom) $T-T_c=6K$, $\delta=0.0095$ and $\zeta=1.42$. Time $t$ is dimensionless (reference time is $\tau_0=\frac{\omega_0}{U_0}$). The inset
shows the interface shape at different times. }\label{dynamics}
\end{center}
\end{figure}

One of the first studies of the dynamics of free surface flow induced by the radiation pressure has been performed by Ostrovskaya et al. \citep{ostrovskaya87}. It consisted in solving the unsteady, linearized momentum and mass conservation equations in cylindrical coordinates with a linearized boundary condition at the free surface. In a recent paper, Wunenburger et al. \citep{wunenburger06b} generalized this approach to a liquid-liquid interface where the liquids have indentical viscosities. In particular, it was shown that for small deformation amplitudes (where $h\lesssim 1$ and $\pi_0 \lesssim 1$), the
interface dynamics is accurately described by a linear theory of
overdamped interfacial waves \citep{wunenburger06b}. For linearized radiation and Laplace pressures (i.e. $\frac{dz}{dr}<<1$), the following expression of the dimensionless time evolution of the interface hump height was predicted to be:\\
\begin{equation}
h_{lin}(t)=z(r=0,t)=\frac{\pi_0 }{4}\int_{0}^{\infty }e^{-k^{2}/8}\frac{1}{1+Bo/k^{2}}%
[1-exp\left({-t\frac{1+Bo/k^{2}}{4}k}\right)]\frac{dk}{k}\label{eq_ht}
\end{equation}
This prediction has proven to be accurate when compared to experimental data over a wide range of fluid properties.\\
Note that the equilibrium solution can be easily deduced when $t \rightarrow \infty$:\\
\begin{equation}
h_{\infty,lin}=\frac{\pi_0}{8} e^{\frac{Bo}{8}}E_{1}(\frac{Bo}{8})
\label{eqlin}
\end{equation}%
where $E_{1}(x)$ is the exponential integral $\int_{x}^{\infty }\frac{e^{-k}%
}{k}dk$.\\
Comparison of numerical predictions with experimental results of the time evolution of the hump height for large beam illumination ($\pi_0>1$) is depicted in figure \ref{dynamics}. Very good agreement on hump dynamics and profiles (inset) is illustrated for two sets of experimental data ($T-T_c=4.5K$ and $T-T_c=6K$) and for different values of the beam illuminations ranging from $\pi_0 \sim 1.5$ to $5.5$.\\
On the same figure, we have represented the time-dependent solution given by the linear model (Eq. \ref{eq_ht}) when the optical radiation pressure and capillary terms are linearized. This linearization strongly underpredicts both the steady hump height (for $T-T_c=4.5K$ and $\pi_0=5.5$, $h_{\infty,lin}=4$ while $h=7.2$) and the time necessary to reach $99\%$ of the steady hump height ($\tau_{99,lin}=40$ while $\tau_{99}=100$). This shows that the complete forms of both the radiation pressure and capillary pressure terms are necessary to yield an accurate description of the interface dynamics.\\

\newpage
\subsubsection{Steady state}
\begin{figure}[h!!!]
\begin{center}
 \includegraphics[scale=0.6]{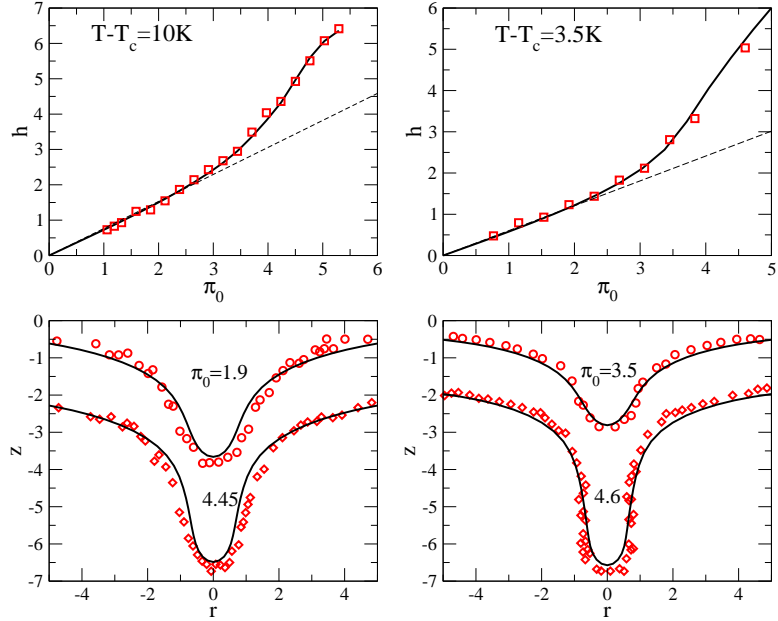}
\caption{(Top) Numerical and experimental interface steady hump height variation with the optical to Laplace pressure ratio $\pi_0$. Dashed lines represent the linear prediction $h_{\infty,lin}$.
(Bottom) Steady state interface profiles for different $\pi_0$ ($\omega_0=5.3 \mu m$) (Profiles were arbitrarily
shifted in the $z$-direction). In all graphs, symbols and continuous
lines represent experimental data and numerical solutions respectively.(Left) $T-T_c=10K$, $\delta=\frac{n_2-n_1}{n_2}=0.011$. (Right) $T-T_c=3.5K$, $\delta=0.008$. }\label{statics}
\end{center}
\end{figure}

In figure \ref{statics}, we compare the experimental results for the stationary interface hump height $h$ versus $\pi_0$ to the computed predictions. These data were obtained at $%
T-T_{c}=10K$ and $T-T_{c}=3.5K$.
Experimental profiles of the interface are also compared to the steady profiles obtained numerically for $\pi_0$ ranging from $2$ to $4.5$. These results clearly show that the numerical resolution
gives an excellent prediction of the interface hump height and profile
providing a quatitative validation of our model in both the linear ($\pi_0 \sim 2$) and
nonlinear ($\pi_0 \sim 4.5$) regime.\\
Considering the accuracy of our calculations, we now extend the numerical investigation to a range of values of the relative index contrast $\delta=\frac{n_2-n_1}{n_2}$ and viscosity ratio $\zeta=\frac{\mu_1}{\mu_2}$ that goes far beyond the existing experimental data. This will highlight original features for both the steady state and dynamics of the interface.\\

\subsection{Optical effects}
The near-critical fluid phases used in the experiments had a relative index contrast varying from $\delta=0.0066$ for $T-T_c=2K$ to $\delta=0.015$ for $T-T_c=25K$ featuring a very narrow range of variation of this parameter. Experimentally, it is possible to reach steady deformations with comparable aspect ratios using other couples of fluids such as Octane/Water with Span 80 or SDS as surfactants ($\sigma \sim 10^{-2}mN/m$ and $\delta \sim 0.06$). Higher relative index contrasts $\delta \sim 0.35$, (Water+Lung surfactants)/Air interface ($\sigma \sim 10mN/m$) can be achieved, altough a beam power of approximately $10W$ becomes necessary.\\
Given these experimental constraints, we investigate the role of $\delta$ in the range $10^{-5} < \delta <0.4$ by means 
of numerical simulations in order to understand its effects on the deformation in the nonlinear regime.\\

\begin{figure}
\begin{center}
\includegraphics[scale=0.8]{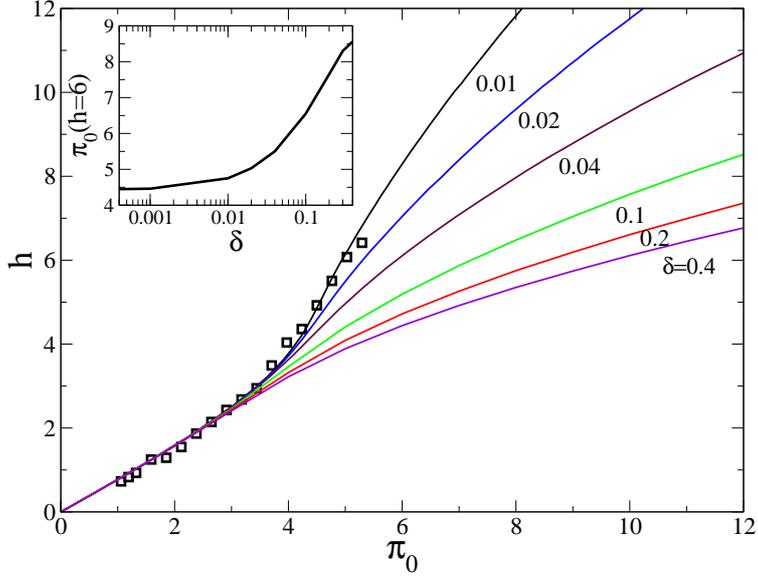}
\end{center}
\caption{Interface hump height variation with $\pi_0$ for different
relative index contrasts $\protect\delta=\frac{n_2-n_1}{n_2}$. The square symbols are experimental data corresponding to $\delta=0.01$. (Inset) Variation of $\pi_0$ needed to reach $h=6$, as a function of $\delta$.}
\label{heights}
\end{figure}

\begin{figure}
\begin{center}
\includegraphics[scale=0.4]{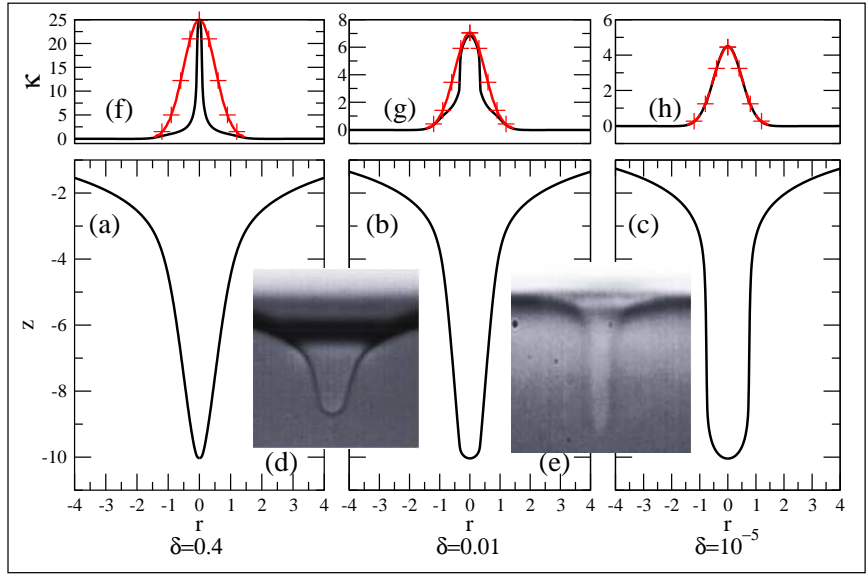}
\end{center}
\caption{Bottom (a-b-c): Interface steady profiles for $\delta=0.4$%
, $0.01$ and $10^{-5}$. $\pi_0$ is chosen to keep $h\simeq10$. $%
Bo=0.003$. Central pictures (d-e) were obtained from experiments with $Bo=0.003$.
Left picture (d): $T-T_{c}=20K$, $\delta \simeq 0.014$, $\pi_0 \sim 5.5$. Right picture (e): $T-T_{c}=2K $, $\delta \simeq 0.0066$ and $%
\pi_0 \sim 9$. Top (f-g-h): Curvature of steady profile associated to the deformations. The line with (+) symbols is a gaussian profile.}
\label{profils}
\end{figure}

In figure \ref{heights}, we have represented the variation of the steady state interface hump height, $h$ as a function of $\pi_0$ for several values of $\delta$ ranging from $0.01$ to $0.4$.\\
As shown above, for sufficiently small values of the optical illumination ($\pi_0<2$), $h$ varies linearly with $\pi_0$ and has no dependence on $\delta$.\\ In the nonlinear regime ($\pi_0>2$), $h$ strongly depends on $\delta$. For any given value of $\pi_0$, $h$ is larger when $\delta$ decreases. This is due to the fact that, for given $\pi_0$ and $\theta_i$, $\Pi(\theta_i$) is a decreasing function of $\delta$ (see figure \ref{pi}).\\
The inset of figure \ref{heights} illustrates this statement. It shows that the value of $\pi_0$ needed to reach $h=6$ as a function of $\delta$ decreases when decreasing $\delta$. However, we can also notice a saturation of $\pi_0=f(\delta)$. Indeed, when $\delta$ is getting very small, the radiation pressure becomes quasi insensitive to the incidence angles $\theta_i$ (i.e. to the local slope $\frac{dz}{dr}$) except when $\theta_i$ is very close to $90°$.\\
\newline
In figure \ref{profils} (a-b-c) we have represented the interface steady
profile, for three values of $\delta $: $10^{-5}$, $0.01$ and $0.4$. The corresponding values of $\pi_0 $ were chosen in order to obtain the same
interface hump height $h\simeq 10$. It is noteworthy that the shape of the deformation in the nonlinear
regime strongly depends on $\delta $. At relatively large values of 
$\delta $, (e.g. $\delta =0.4$ in figure \ref{profils}a), the center part of the
interface adopts a needle-like shape with a sharp tip.\ Decreasing $\delta $ (e.g. $%
\delta =0.1$, see figure \ref{profils}b) widens and rounds off the tip, showing a bell shape, while
the base of the deformation shrinks.\ More surprisingly, when the two fluids
have almost the same refractive properties, a stable finger, nearly
cylindrical with a round tip, is obtained (figure \ref{profils}c). This shape evolution corresponds to what is observed on the two
pictures superimposed on figure \ref{profils} and extracted from experiments performed with near-critical phases.
 Figure \ref{profils}d, obtained at $%
T-T_{c}=20K$ ($\delta \simeq 0.014$) and $\pi_0 \sim 5.5$ shows that the
interface adopts a bell\ shape while figure \ref{profils}e, obtained at $%
T-T_{c}=2K $ ($\delta \simeq 0.0066$) and $\pi_0 \sim 9$, shows a nearly
cylindrical finger.\\ These observations can be explained qualitatively by considering the balance of radiation and Laplace pressures. The finger shape ($\delta \rightarrow 0$) corresponds to a case where the radiation pressure is almost independent of the incident angle (see figure \ref{pi}). This means that $\Pi(r) \simeq \pi_0 e^{-2r^2}$. Therefore, as only the capillary force balances the optical radiation pressure at steady state (we neglect the weak influence of gravity), we have $\kappa(r) \sim \Pi(r)$. This is confirmed on figure \ref{profils}-h where the curvature profile shows an excellent agreement with the gaussian profile (see figure \ref{profils}h).\\ The other shapes ($\delta =0.1$ and $\delta =0.4$) show that when the radiation pressure strongly depends on the incidence angle, the profile tends to a needle shape as the maximum of $\Pi(r)$ becomes more and more localized toward the tip ($r=0$ and $\theta_i=0$).\\
Consequently, one can tune the morphology of stationary nonlinear deformations by simply changing the refractive index contrast.\\
In the following, we show that the dynamics of the deformation has also a strong dependence on the relative index contrast $\delta$.\\

\begin{figure}
\begin{center}
\includegraphics[scale=0.6]{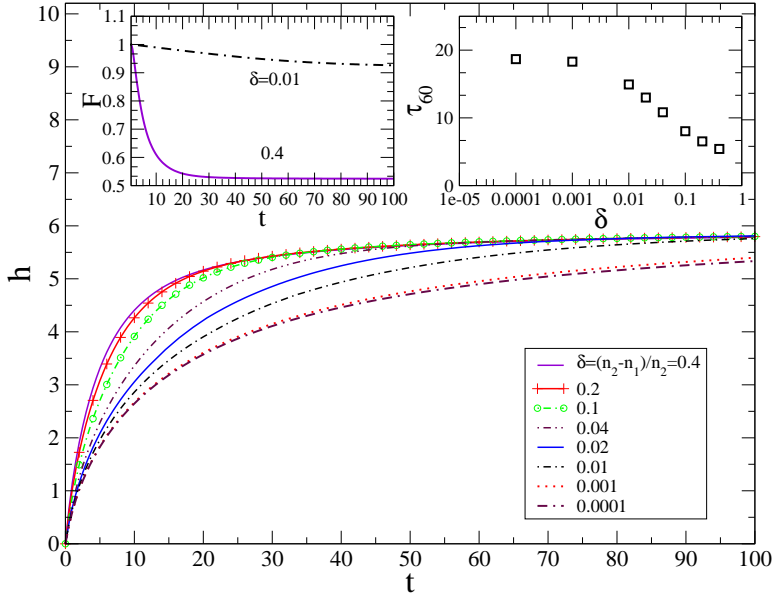}
\end{center}
\caption{Time evolution of the interface hump height for different values of $\delta$. $\pi_0$ is chosen to keep $h\simeq6$. $Bo=0.005$. The left inset shows the time variation of the net radiation force acting on the interface $F(t)$ for $\delta=0.4$ and $\delta=0.01$ ($\zeta=\frac{\mu_1}{\mu_2}=1$). The right inset shows the characteristic time $\tau_{60}$ taken for the hump height to reach $60\%$ of its final steady value plotted against $\delta$.}
\label{dyn_index}
\end{figure}

In figure \ref{dyn_index}, the time evolution of the hump height $h(t)$ is plotted
for $\delta$ ranging from $10^{-4}$ to $0.4$. 
We note that the transient regime is significantly longer when $\delta$ is closer to zero, i.e. when the fluids tend to match optically. This is explained in the left inset of this figure that shows the time variation of the radiation pressure integrated over the interface $|\int_{S_I} \Pi(r,t){\bf n}dS|=2\pi\int_0^{\frac{R}{\omega_0}} \Pi(r,t)rdr$ reduced by its initial value $2\pi\int_0^{\frac{R}{\omega_0}} \Pi(r,t=0^+)rdr$. This ratio, called $F$, simply represents the time variation of the net optical force applied to the interface during the transient deformation. We can notice that this force quickly reaches its steady state value when $\delta=0.4$ whereas the transient is much longer for $\delta=0.01$. This observation can be explained qualitatively by considering the decrease of the radiation
pressure with the incidence angle $\theta_i$. As the deformation increases, $\theta_i$ increases along the interface, therefore the associated radiation pressure applying on the growing hump decreases locally for $\delta=0.4$ while it remains quasi unchanged for $\delta=0.01$. The characteristic relaxation time of the optical force is smaller when increasing $\delta$. As this force is the source of momentum transfered to the interface, the dynamics of the interface follows a similar behaviour with $\delta$.\\
The right inset shows a characteristic time $\tau_{60}$ chosen arbitrarely at $60\%$ of the equilibrium hump height as a function of $\delta$. We can clearly see that $\tau_{60}$ increases when $\delta$ decreases, before reaching a saturation. This saturation is related to the weak dependence of $\Pi(r)$ on $\theta_i$ at small $\delta$.\\ 

\subsection{Viscous effects}

By analogy with the analysis of the effects of refraction indices on the deformation, we now investigate the effect of the viscous ratio $\zeta=\frac{\mu_1}{\mu_2}$ on the transient behaviour of the deformation towards equilibrium. Equation (\ref{stressjump}) shows that the transient deformation depends on viscosities only through their ratio $\zeta$.\\
In the case of the near-critical phases used in the experiments reported earlier, $\zeta$ was varying from $1.27$ for $T-T_c=2K$ to $\zeta=1.76$ for $T-T_c=25K$. As $\zeta$ was close to unity, no particular behaviour was observed in the transient regime. \\
It is however possible to imagine experiments with larger values of $\zeta$. Using viscous Glucose-Water/Hexadecane or Glucose-Water/Toluene and appropriate surfactants would enable to obtain a viscosity ratio of $\zeta \sim 150$ with $\delta \sim 0.06$ and $\sigma\sim  10^{-2}mN/m$ still allowing large scale deformations using a continuous laser beam. A theoretical study would thus be a very useful predictive tool for the design of realistic future experiments. For this reason, we investigate the effect of the viscosity ratio in the two limit cases $\zeta=200$ and $\zeta=1/200$.\\

\begin{figure}
\begin{center}
\includegraphics[scale=0.65]{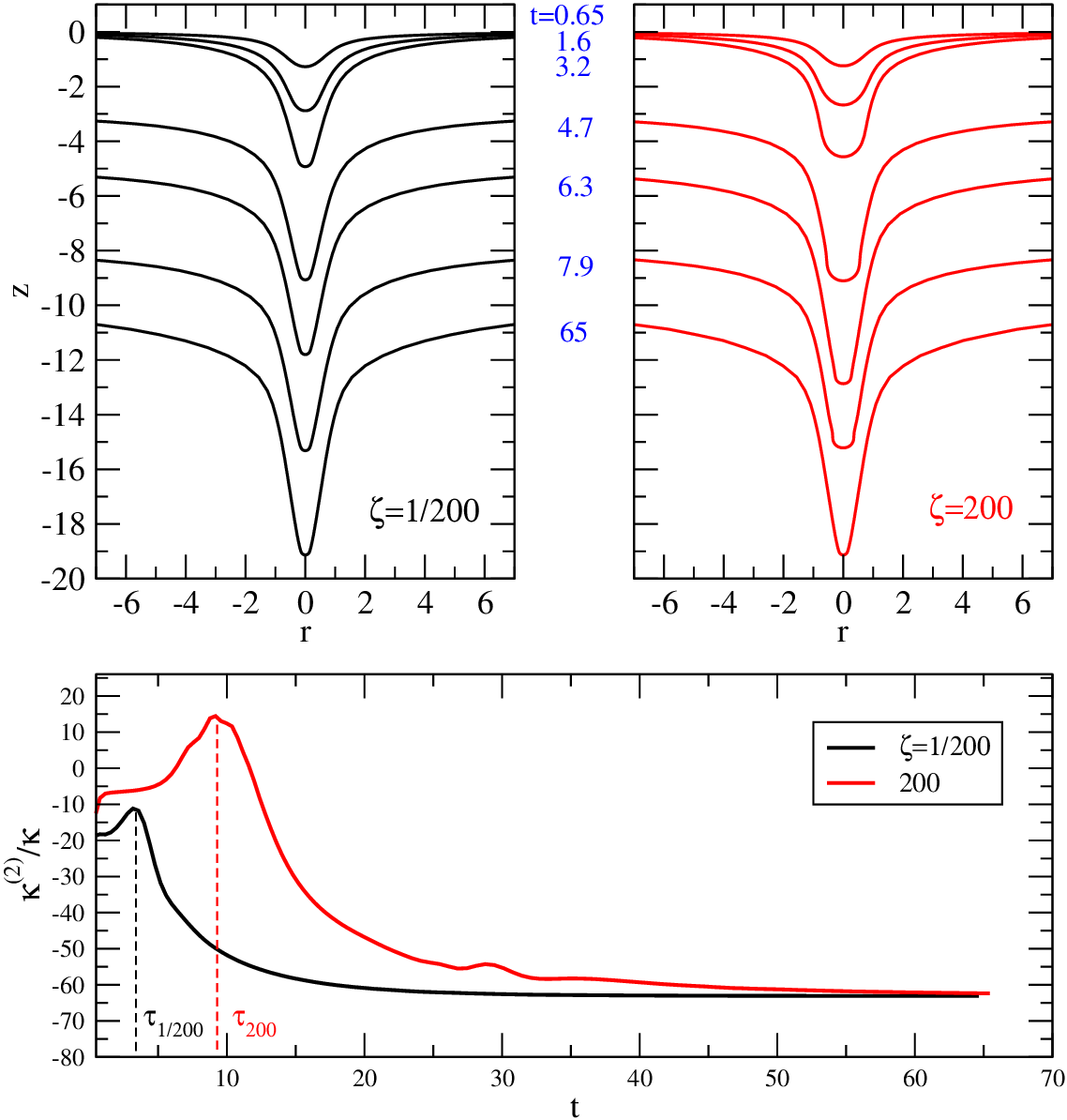}
\end{center}
\caption{(Top) Interface profiles at different times for two values of the viscosity ratio $\zeta=\frac{\mu_1}{\mu_2}=\frac{1}{200}$ (left) and $\zeta=200$ (right). $\delta=0.1$ and $Bo=0.01$. Profiles were shifted vertically for clarity. (Bottom) Time evolutions of the corresponding reduced curvature of the curvature $\displaystyle{\frac{\kappa^{(2)}(t)}{\kappa(t)}=\frac{\frac{d^ 2\kappa(r,t)}{dr^2}|_{r=0}}{\frac{d^ 2z}{dr^ 2}|_{r=0}}}$.}
\label{viscous}
\end{figure}

Figure \ref{viscous} shows the evolution of the interface profile at different times for these two values of $\zeta$, keeping $\delta=0.1$, $Bo=0.01$ and $\pi_0=15$.\\ In the case $\zeta=\frac{1}{200}$, we can observe a sharp transition from a bell shape ($t=0.65$), characteristic of a small deformation amplitude, to a needle shape ($t=4.7$). For $\zeta=200$, we observe that the transition towards the needle shape occurs later, the tip of the interface showing a rounded shape between $t=4.7$ and $t=7.9$. This shape transition can be further characterized by analyzing the time dependence of the reduced curvature of the curvature at the tip, $\displaystyle{\frac{\kappa^{(2)}(t)}{\kappa(t)}=\frac{\frac{1}{r}\frac{d}{dr}\left(r\frac{d\kappa(r,t)}{dr}\right)|_{r=0}}{\kappa(r=0,t)}=\frac{\frac{d^ 2\kappa(r,t)}{dr^2}|_{r=0}}{\frac{d^ 2z}{dr^ 2}|_{r=0}}}$ shown in figure \ref{viscous}, 
$\frac{\kappa^{(2)}}{\kappa}$ being an indicator of the spatial variation of the curvature. In fact, a needle shape has a sharper curvature variation with $r$ than a rounded shape (see figure \ref{profils}(f) and (h)), which means that $\frac{\kappa^{(2)}}{\kappa}$ is larger (in magnitude) for the needle shape.
We observe in figure \ref{viscous} that for a given value of $\zeta$, $\frac{\kappa^{(2)}(t)}{\kappa(t)}$ initially increases until reaching a maximum at time $\tau_\zeta$ ($\tau_{1/200}=3.3 $, $\tau_{200}=9.5 $) before decreasing toward the steady state value . This maximum is a signature of the change of the shape evolution that switches from a round tip to a sharp one corresponding to a needle shape.\\

\section{Conclusion}
\begin{figure}
\begin{center}
\includegraphics[scale=0.5]{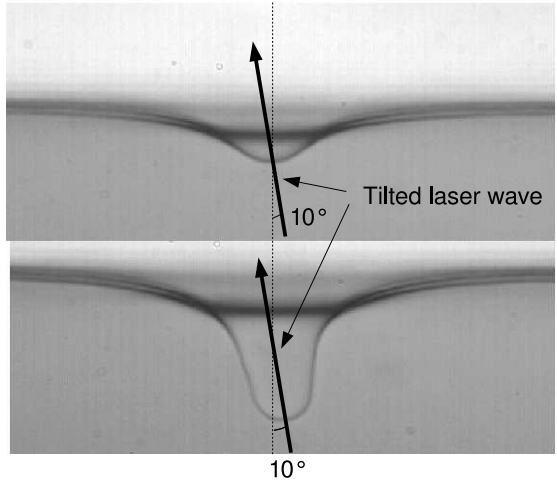}
\end{center}
\caption{Interface deformation with a tilted laser beam propagating upwards. Top : $P=1W $. Bottom $P=1.7W$. $n_1-n_2=0.016$ and $\sigma=1.4~10^ {-6}N/m $. When the beam is tilted, incidence angles on one side of the deformation (here the left one) are larger than in the other side. When they are sufficiently small to consider radiation pressure as constant all over the interface, the deformation remains symmetric (Top picture). When the beam power is increased, the growth of the deformation makes radiation pressure sensitive to incidence angles and then more efficient on the right than on the left. The deformation is thus attracted in the beam propagation direction and becomes non-axisymmetric (Bottom picture).}
\label{oblique}
\end{figure}

While quantitatively validated with experimental data in the nonlinear regime of deformation, our numerical resolution, based on a Boundary Element Method, showed that new morphologies of a soft fluid-fluid interface deformed by a continuous laser wave emerge when fluids become contrasted.
In the nonlinear regime, where the radiation pressure and the height of the interface strongly depends on the relative refractive index contrast $\delta$, we showed that the interface shape turns from a needle to a nearly-cylindrical finger shape.
In the transient regime, we predicted that the characteristic time of the deformation increases when decreasing $\delta$ before reaching a saturation. These results show the strong nonlinear coupling between the radiation pressure and the interface deformation in both transient and steady state. It was shown that the physical feature explaining these results lies in the dependence of the radiation pressure on the incidence angle. An original experimental evidence of this dependence can be achieved when the beam axis is not perpendicular to the initial flat interface. In this configuration, the interface hump is attracted in the direction of light propagation in the nonlinear regime of deformation (see figure \ref{oblique}).\\
Finally, the influence of the viscosity ratio $\zeta$ has been investigated for large scale deformations. We showed that the transition time of the interface from its initial bell shape to the steady shape increases with $\zeta$.\\
All these results show that the morphology of nonlinear interface deformations driven by the radiation pressure of a laser wave is even more rich than expected, opening the route towards future original experiments. Further developments on nonlinear effects, where a feedback coupling between deformation and propagation emerges, will extend deformation shapes to self-adapted ones as for liquid optical fibres \citep{brasselet08} and nipple-like shapes observed experimentally \citep{casner03a}. They both involve light guiding within the deformation that modifies in turn the radiation pressure along the structure. This thorough analysis of optohydrodynamics provides the general frame to predict and anticipate further developments of contactless interface manipulation at the micrometer scale.

~~~~~~~~~~~~~~~~\\
\textbf{ Acknowledgements}\\

This research was supported by Centre National de la Recherche
Scientifique (France), Universit\'e Bordeaux 1, and Conseil
R\'egional d'Aquitaine (Contract $N° 20051101010A$).
We thank Julien Petit for his contribution to Figure \ref{oblique}.
%\bibliographystyle{unsrt}
%\bibliography{hamza_min}

\end{document}